\documentclass[doublecol]{epl2} 

\usepackage[bookmarks=true, colorlinks=true, linkcolor=blue, urlcolor=black, citecolor=blue, pdftex, hyperindex=true]{hyperref}
\usepackage{dcolumn}
\usepackage[normalem]{ulem}
\usepackage{amsmath} 
\usepackage{amssymb}
\usepackage{amsfonts}
\usepackage{bm}
\usepackage[dvips]{epsfig}
\usepackage{color}
\usepackage[dvipsnames]{xcolor}
\usepackage{comment}
\usepackage{mathtools} 
\usepackage{lipsum}
\usepackage{etoolbox}
\usepackage{graphicx}
\usepackage{cuted}
\usepackage{caption}
\usepackage{ragged2e}

\newcommand{\cs}{c_{\mbox{\scriptsize s}}}
\newcommand{\taulb}{\tau_{\mbox{\tiny LB}}}
\newcommand{\area}{\chi}
\renewcommand{\vec}[1]{\boldsymbol{\mathbf{#1}}}

\newcommand{\Rey}{\text{Re}}
\newcommand{\Ca}{\text{Ca}}

\title{On the hydrodynamic behaviour of the immersed boundary - lattice Boltzmann method for wetting problems}
\shorttitle{Hydrodynamic behaviour of IBLB method for wetting problems} 

\author{E. Bellantoni\inst{1,2,3} \and F. Guglietta\inst{2} \and A. Demou\inst{1} \and F. Pelusi\inst{4} \and K. Um\inst{3} \and M. Nicolaou\inst{1,6} \and M. Desbrun\inst{5} \and M. Sbragaglia\inst{2} \and N. Savva\inst{1,7} }
\shortauthor{E. Bellantoni \etal}

\institute{   
  \inst{1}CaSToRC, The Cyprus Institute, 20 Konstantinou Kavafi Street, 2121 Nicosia, Cyprus\\
  \inst{2}Dept.~of Physics and INFN, Tor Vergata University of Rome, Via della Ricerca Scientifica, 1-00133 Rome, Italy\\
  \inst{3}LTCI, T\'{e}l\'{e}com Paris, IP Paris, 19 Place Marguerite Perey, 91120 Palaiseau, France\\
  \inst{4}CNR - Istituto per le Applicazioni del Calcolo, Via dei Taurini 19, 00185 Rome, Italy\\
  \inst{5} Inria and LIX (\'Ecole Polytechnique), IP Paris, 1 rue Honor\'e d'Estienne d'Orves, 91120 Palaiseau, France\\
  \inst{6}Dept.~of Computer Science, University of Cyprus, 1 Panepistimiou Avenue, 2109 Nicosia, Cyprus \\
  \inst{7}Dept.~of Mathematics and Statistics, University of Cyprus, 1 Panepistimiou Avenue, 2109 Nicosia, Cyprus
 }
\abstract{
We study the hydrodynamic behaviour of a mesoscale numerical model for wetting dynamics based on the immersed boundary - lattice Boltzmann (IBLB) method. This IBLB model features a wetting potential to capture the interaction between a non-ideal droplet interface and a solid boundary; it is designed to prevent abrupt curvature changes near the contact line.
As this approach prevents direct contact between the droplet and the solid, it forms a thin film beneath the droplet, which could compromise the hydrodynamic consistency in this region. 
This paper presents detailed comparisons against two other hydrodynamic solvers, respectively based on a boundary element method (BEM) and a volume of fluid (VoF) method, in order to examine the hydrodynamic behaviour of this IBLB scheme, elucidate its limits of validity in wetting applications, and explore the properties of its contact-line model.
}
\begin{document}
\maketitle
\section{Introduction}\label{sec:introduction}
Wetting is a complex process whose study intertwines physics, chemistry, and engineering~\cite{DeGennes1985,DeGennes2004capillarity,bonn2009wetting}. To uncover the multiscale mechanisms driving the spreading of a liquid droplet over a solid surface, a vast amount of effort has been invested across experiments, theory, and numerical simulations, as highlighted by reviews on the topic~\cite{DeConinck2008,snoeijer2013,andreotti2020}.
Over the past several decades, numerical simulations have become a traditional means to describe and investigate wetting~\cite{deruijter1999dynamic,carlson2011dissipation,kusumaatmaja2006drop,sbragaglia2007spontaneous,ding2007inertial,savva2010two,wheeler2010modeling,du2021initial,legendre2013}.
Among these, the lattice Boltzmann (LB) method has established itself as a versatile tool. By leveraging mesoscale modeling, the LB method accurately recovers the hydrodynamic behavior of a system - provided there is a clear separation of scales (i.e., in the Chapman-Enskog limit)~\cite{kruger2017,succi2018}. Since its introduction, numerous developments to the LB method have emerged, including the possibility of modeling non-ideal fluids~\cite{Swift1995,Shan1993,Shan1994}. Beyond traditional diffuse interface methods (see Refs.~\cite{kruger2017,succi2018} for a review on the topic), other approaches have emerged, such as the immersed boundary - lattice Boltzmann (IBLB)~\cite{kruger2011} method. 
The IBLB method has been applied to a series of problems involving the motion of complex interfaces in viscous fluids~\cite{kruger2011,guglietta2020effects,taglienti2024}, including recent applications to wetting problems~\cite{pelusi2023sharp,bellantoni2025immersed}. In Ref.~\cite{bellantoni2025immersed}, an IBLB method was proposed using a disjoining-pressure-like contribution to regularize the droplet interface curvature near flat solid boundaries. This interaction term prevents direct contact between the droplet and the solid, leading to the formation of a thin film beneath the droplet. The benchmarks examined in Ref.~\cite{bellantoni2025immersed} addressed the correctness of the droplet's shape at equilibrium and the recovery of known scaling exponents for spreading dynamics on homogeneous surfaces. However, the thin film existing near the contact line raises questions regarding the consistency of the IBLB method in reproducing the correct hydrodynamics in this region. 
Furthermore, Ref.~\cite{bellantoni2025immersed} examined only the steady-state shapes of droplets against theoretical predictions, making an investigation of shapes during spreading a natural complement.
To address these points, a detailed benchmarking against alternative hydrodynamic solvers is required to confirm that the IBLB model presented in Ref.~\cite{bellantoni2025immersed} accurately recovers the expected hydrodynamic limit and droplet's shape evolution during wetting dynamics.
This need motivates the present contribution. 
\begin{figure*}[t]
    \centering
    \includegraphics[width=\linewidth]{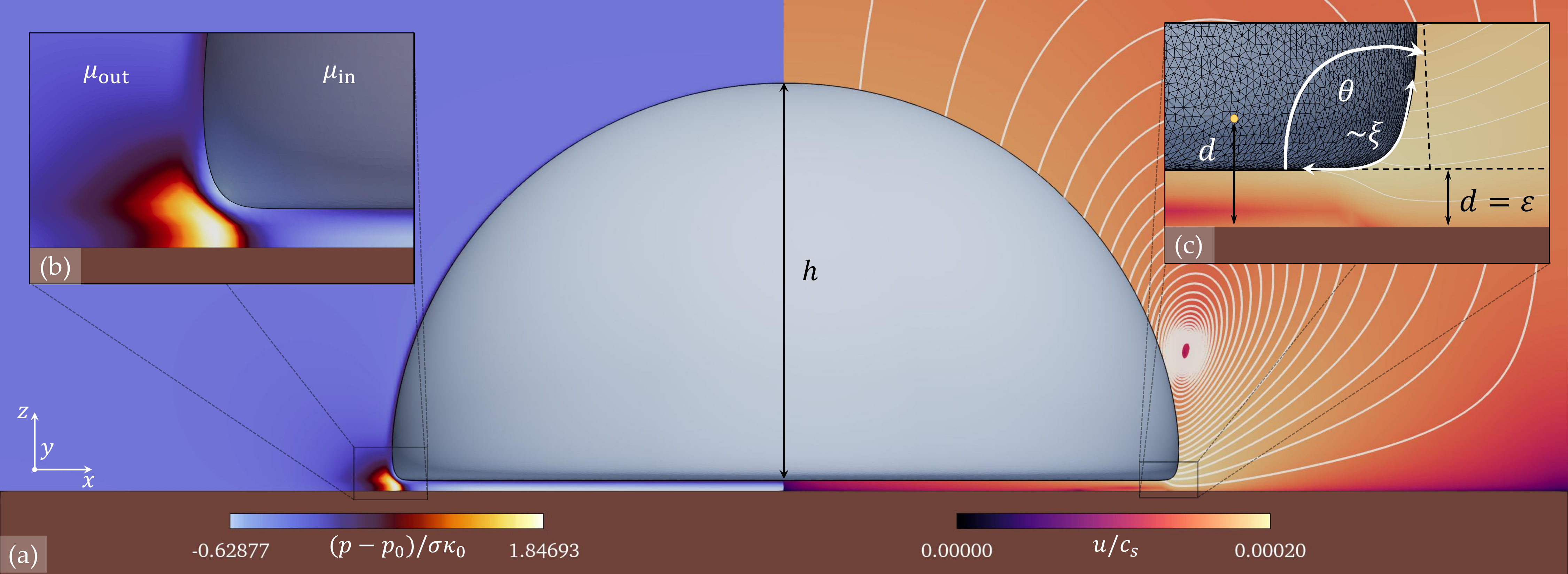}
    \caption{
    Split view of the hydrodynamic fields around a droplet wetting a solid, simulated with the IBLB method (see triangular mesh in inset (c)). On the left, the pressure field on a slice passing through the center of the droplet is shown as the difference from the initial homogeneous pressure $p_0$ in Laplace pressure units $\sigma \kappa_0$, where $\sigma$ is the surface tension and $\kappa_0$ is the droplet's curvature at its initial spherical configuration; an extra pressure develops near the contact-line region due to the thin film presence (see inset (b)). On the right, the velocity magnitude on a slice passing through the center of the droplet is shown as a color map in units of the lattice sound speed $c_s$, together with streamlines (white lines) of the velocity field in the $xz$-plane. 
    \vspace*{-2mm}
    }
    \label{fig:choreo}
\end{figure*}
\section{Wetting via immersed boundary - lattice Boltzmann method}
In this section, we review the IBLB method for wetting problems presented in Ref.~\cite{bellantoni2025immersed}. Based on discrete kinetic theory, the LB method tracks the evolution of the probability distribution function (PDF), $f_i(\vec{x},t)$. It describes the likelihood of a particle residing at position $\vec{x}$ on a three-dimensional Cartesian lattice (with spacing $\Delta x$) at time $t$, moving with a discrete kinetic velocity $\vec{c}_i$ from a finite set ($i=0,\dots,Q-1$); in this work, we employ a D3Q19 velocity stencil, meaning that we work in three dimensions with $Q=19$ velocity vectors.
The dynamics of $f_i(\vec{x},t)$ is set by the {\it lattice Boltzmann equation}, encoding both streaming and relaxation over a time lapse $\Delta t$~\cite{kruger2017,succi2018}:
\begin{equation}\label{eq:LBE}
f_i(\vec{x}+ \vec{c}_i \Delta t, t + \Delta t) -f_i(\vec{x}, t) = \Delta t \left[ \Omega_i(\vec{x},t) + S_i(\vec{x},t) \right]\!.
\end{equation}
The streaming step on the left-hand side of Eq.~\eqref{eq:LBE} is exact on the lattice: the PDF is advected from one node to its neighbor along the discrete velocity vector $\vec{c}_i$. The term $\Omega_i(\vec{x},t)$ represents the relaxation process, expressed by the Bhatnagar-Gross-Krook (BGK) collision operator~\cite{bhatnagar1954}:
\begin{equation}
\Omega_i(\vec{x},t) = - \frac{1}{\taulb}\left[ f_i(\vec{x}, t) - f_{i}^{(\rm eq)}(\vec{x}, t) \right].
\end{equation}
This collision model features a characteristic relaxation time $\taulb$ regulating the relaxation of the PDF towards an equilibrium distribution $f_{i}^{(\rm eq)}(\vec{x},t)=f_{i}^{(\rm eq)}(\rho(\vec{x},t),\vec{u}(\vec{x},t))$ that depends on the fluid density, $\rho=\rho(\vec{x},t)$, and velocity, $\vec{u}=\vec{u}(\vec{x},t)$, via a truncated Taylor expansion of the Maxwell-Boltzmann PDF~\cite{kruger2017,succi2018}:
\begin{equation}
\small
f_{i}^{(\rm eq)}(\rho, \vec{u}) = \omega_i \rho \left[ 1 + \dfrac{ \vec{u} \cdot \vec{c}_i}{ \cs^2 } + \dfrac{ (\vec{u} \cdot \vec{c}_i)^2}{ 2\cs^4 } - \dfrac{ \vec{u} \cdot \vec{u}}{2 \cs^2 } \right], 
\end{equation}
where $\cs=\Delta x / (\sqrt{3}\Delta t)$ and $\omega_i$ are statistical weights associated with the velocities $\vec{c}_i$. The source term $S_i(\vec{x},t)=S(\vec{u}(\vec{x},t),\vec{F}(\vec{x},t))$ appearing in Eq.~\eqref{eq:LBE} depends on both velocity and force density $\vec{F}=\vec{F}(\vec{x},t)$ as~\cite{guo2002} 
\begin{equation}\label{eq:source}
\small
S_i(\vec{u},\vec{F}) = \left( 1 - \frac{\Delta t}{2 \taulb} \right) \omega_i \left[ \dfrac{\vec{c}_i - \vec{u}}{ \cs^2 } + \left(\dfrac{ \vec{u} \cdot \vec{c}_i}{ \cs^4 }\right) \vec{c}_i \right] \cdot \vec{F}\ .
\end{equation}
The fluid density $\rho$ and velocity $\vec{u}$ are defined in terms of the first two moments of the PDF~\cite{guo2002,kruger2017}, that is,
\begin{gather}
\rho = \sum_i f_i \label{eq:hydro_var1} \, , \\ 
\vec{u} = \frac{1}{\rho}\sum_i \vec{c}_i f_i + \dfrac{\vec{F} \Delta t}{2\rho} \, ,\label{eq:hydro_var2}
\end{gather}
where Eq.~\eqref{eq:hydro_var2} contains the half-force correction~\cite{guo2002,kruger2017}.\\ 
The interface is made up of $N$ Lagrangian nodes with positions $\vec{q}_{j}(t)$ ($j=1,\dots, N$) arranged in a 3D triangular mesh. The number of triangles is kept fixed throughout the system's evolution, and the IB technique~\cite{Peskin1972,peskin2002} is used to handle fluid-structure interaction. 
A two-way coupling governs this interaction: interface nodes are updated via a forward Euler scheme, $\vec{q}_j(t + \Delta t) = \vec{q}_j(t) + \dot{\vec{q}}_j(t) \Delta t$; Lagrangian velocities $\dot{\vec{q}}_j$ are interpolated from the fluid to satisfy the no-slip condition, while the back-reaction is captured by spreading interface nodal forces $\vec{\varphi}_j$ onto the fluid grid to determine the local force density $\mathbf{F}$ in Eq.~\eqref{eq:source}:
\begin{gather}\label{eq:vel_interpolation}
    \dot{\vec{q}}_j(t) = \sum_{\vec{x}} \vec{u}(\vec{x},t) \Delta(\vec{q}_j(t) - \vec{x})\Delta x^3 ,  \\
\label{eq:force_spread}
    \vec{F}(\vec{x},t) = \sum_{j} \vec{\varphi}_j(t) \Delta(\vec{q}_j(t)-\vec{x}) ,
\end{gather}
where $\Delta(\vec{x})$ is a discrete Dirac delta function constructed with interpolation stencils~\cite{kruger2017,peskin2002}.\\ 
The force density $\vec{F}$ is a crucial ingredient in the IBLB method, as it enables the modeling of non-ideal interface properties that include surface tension and wall interaction terms. The value of $\vec{F}$ around a node position $\vec{q}$ is the sum of a surface tension term with surface tension $\sigma$, and a wetting contribution with interaction term $\Pi(d)$:
\begin{equation}\label{eq:force_split}
\vec{F}(\vec{x},t) = -\left[\sigma \vec{\hat{n}} \, (\vec{\nabla} \cdot \vec{\hat{n}}) + \Pi(d) \vec{\hat{n}} \right]\delta(\vec{x}-\vec{q})\ ,
\end{equation}
where $\vec{\hat{n}}$ is the normal at the interface pointing outside the droplet and $\delta(\vec{x})$ is the Dirac delta function. Specifically, the implementation of the surface tension term for a clean droplet with a given $\sigma$ is achieved via a finite-element implementation of the strain energy~\cite{pelusi2023sharp,green1960large,barthes1981time,dimitrienko2010nonlinear}, while the wetting interaction term adopted is~\cite{bellantoni2025immersed,chamakos2016droplet,du2021initial}
\begin{equation}\label{eq:disjoin-p}
\Pi(d) = A\left[\left( \frac{\xi}{d}\right)^n-\left( \frac{\xi}{d}\right)^m\right] ,
\end{equation}
where $n$ and $m$ ($n>m$) are parameters regulating the range of the interaction, $A$ is a constant, and $d$ measures the (vertical) distance from the solid surface (see Fig.~\ref{fig:choreo}). The parameter $\xi$ is a {\it regularization lengthscale} that allows for smooth curvature changes close to the wall, with $\xi \rightarrow 0$ resulting in more abrupt curvature changes. Within this approach, a film of thickness $\varepsilon$ forms below the droplet, with the property that $\varepsilon\to 0$ as $\xi\to 0$. The wetting interaction term is implemented via the nodal forces $\vec{\varphi}_{{\scriptscriptstyle\Pi},_j} = - \area_j\Pi(d) \vec{\hat{n}}_j/3$, 
where $\vec{\hat{n}}_j$ is the average of the normal vectors of all faces that share the $j$-th node, while $\area_j$ represents the 1-ring area of the mesh associated with the $j$-th node. The factor $1/3$ accounts for each triangle area being shared by three vertices. 
The parameters $m$, $n$, $A$, and $\xi$ of the wetting-interaction force can be linked to the static contact angle. In Ref.~\cite{bellantoni2025immersed} it was shown that, when the static interface profile is close to a wedge-like structure, the wedge displays an equilibrium contact angle $\theta_{\rm eq}$, which is related to $\sigma$, $A$, $n$, and $m$ as follows~\cite{chamakos2016droplet,karapetsas2016,du2021initial}:
\begin{equation}\label{eq:A_choice}
A =\sigma \frac{(m-1)(n-1)}{(n-m)\xi}(1+\cos\theta_{\rm eq}).
\end{equation}
At the macroscale, the continuity and the Navier-Stokes equations (NSE) are recovered for the fluid density and velocity in Eqs.~\eqref{eq:hydro_var1}-\eqref{eq:hydro_var2}:
\begin{gather}
\dfrac{\partial \rho}{\partial t} + \vec{\nabla}\cdot (\rho \vec{u})=0 \ , \label{eq:continuity} \\
\rho \left[ \dfrac{\partial \vec{u}}{\partial t} + (\vec{u} \cdot \vec{\nabla}) \vec{u} \right] = - \vec{\nabla}{p} + \mu \nabla^2{\vec{u}} + \vec{F} \ , \label{eq:NSE}
\end{gather}
where $p=p(\vec{x},t)=\cs^2\rho(\vec{x},t)$ is the fluid pressure, and $\mu$ is the dynamic viscosity set by $\mu =\rho \cs^2 (\taulb - \Delta t /2)$. To account for the varying fluid viscosity inside and outside the droplet, we consider the LB relaxation time as a function of both space and time, $\taulb=\taulb(\vec{x},t)$: using a ray tracing algorithm, we identify fluid nodes inside and outside the droplet and assign them two different relaxation times, $\taulb^{\text{in}}$ and $\taulb^{\text{out}}$, respectively. This results in a viscosity ratio between the inner and outer viscosity $\lambda=\mu_{\text{in}}/\mu_{\text{out}}$, where $\mu_{\text{in}}=\rho \cs^2 (\taulb^{\text{in}} - \Delta t /2)$ and $\mu_{\text{out}}=\rho \cs^2 (\taulb^{\text{out}} - \Delta t /2)$.
\\
The recovery of hydrodynamic equations~\eqref{eq:continuity}-\eqref{eq:NSE} within the IBLB framework relies on a clear separation between mesoscopic and hydrodynamic scales. This assumption becomes questionable in the near-wall region, particularly in the limit $\xi \to 0$, where the thin film forming beneath the droplet is resolved by only a few lattice nodes. Therefore, the validity of the hydrodynamic limit must be carefully assessed.
In Ref.~\cite{bellantoni2025immersed}, it was shown that the static solutions (i.e., $\vec{u}=\vec{0}$) recovered from the hydrodynamic equations are well reproduced by the IBLB simulations; furthermore, well-known scaling laws for spreading dynamics are also recovered~\cite{bonn2009wetting,winkels2012,legendre2013}. Yet, comparing the IBLB method to other solvers is highly desirable, both to further strengthen the validity of the IBLB and to highlight its properties in comparison to other contact-line models.
\section{Comparison with boundary element method}\label{sec:BEM}
As a first benchmark, we compare the predictions of our IBLB method with those of the boundary element method (BEM), which is applicable in the Stokes regime. In the BEM, the linearity of the governing equations is exploited to represent the velocity field as a superposition of fundamental singular solutions (i.e., Green's functions), thereby converting the volume partial differential equations into boundary integral equations defined only on the boundary of the evolving droplet~\cite{pozrikidis1990}. Typically, this boundary integral formalism expresses the interfacial velocity as a superposition of two terms: (i) the so-called single-layer potential, which is associated with surface traction (here $\bar{\vec{F}}=\left[\sigma (\vec{\nabla} \cdot \vec{\hat{n}}) + \Pi(d)  \right]\vec{\hat{n}}$), and (ii) a double-layer potential associated with velocity jumps. In the limit of equal viscosities (i.e., $\lambda=1$), the double-layer term vanishes, which leads to a pure single-layer formulation, allowing to explicitly determine the interfacial velocities in terms of a weakly singular surface integral~\cite{Pozrikidis_1992}.

For the axisymmetric geometry considered in this work, the aforementioned surface integral further reduces to a line integral along the meridional generating curve $C$, with the azimuthal dependence integrated analytically, that is,
\begin{equation}
    \vec{U}_C(\vec{x},t)= - \int_C \vec{G}(\vec{x},\vec{y})\bar{\vec{F}}(\vec{y},t)
    \,d s(\vec{y}),\label{eq:bem}
\end{equation}
where $\vec{U}_C$ is the interfacial velocity for $\vec{x}$ on $C$, and $\vec{G}$ is the Stokeslet (i.e., the fluid velocity field generated by a point force), which is specially developed for wall-bounded flows that satisfy the no-slip condition~\cite{Blake1971}. Hence, $\bar{\vec{F}}$ can be evaluated for a given surface shape, which, in turn, determines $\vec{U}_C(\vec{x},t)$ in order to evolve the generating curve $C$.  The latter is discretized using customized cubic splines such that the pertinent symmetry conditions are enforced. The numerical implementation of Eq.~\eqref{eq:bem} proceeds by evolving the interface with an explicit second-order Runge--Kutta method and adaptively refining the nodes on $C$ to accurately resolve high-curvature regions.

The method described above is used to extract the evolution of the height $h(t)$ of a droplet (cf. Fig.~\ref{fig:choreo}), initialized as a spherical cap of radius $R_0$.
\begin{figure}[t!]
    \centering
    \includegraphics[width=\columnwidth]{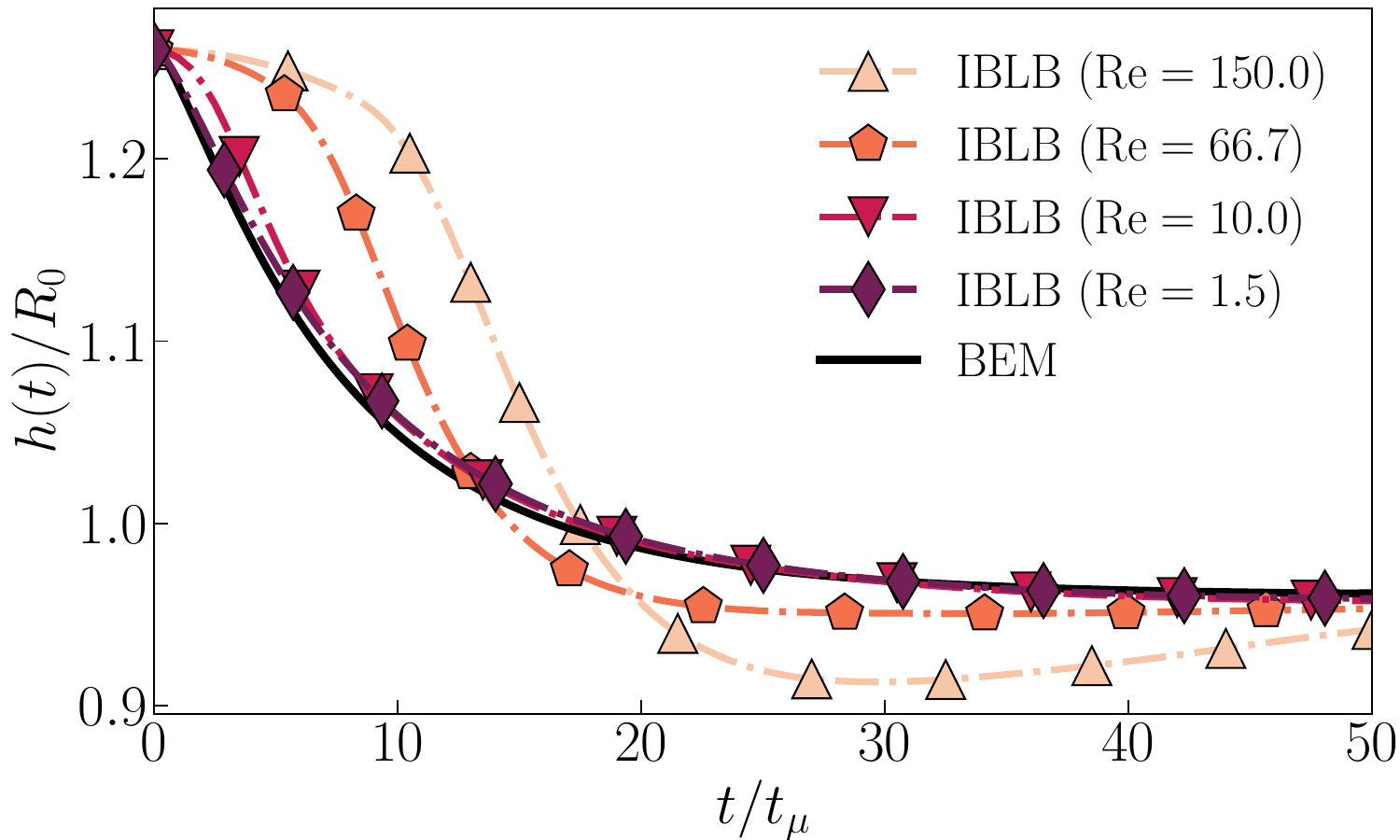}
    \caption{
    Droplet's height evolution during spreading with the BEM and the IBLB method. Dash-dotted lines with different colors/symbols represent IBLB simulation data for different values of the Reynolds number, while the solid black line marks results for a BEM simulation.} \vspace*{-2mm}
    \label{fig:IBLB-BEM}
\end{figure}
In Fig.~\ref{fig:IBLB-BEM}, we consider a hydrophilic spreading process and we compare $h(t)$ resulting from IBLB simulations with the the corresponding the BEM solution. The initial IBLB configuration is an equilibrated droplet with a contact angle $\theta_{\rm eq}=\pi/2$, which spreads to $\theta_{\rm eq}=\pi/3$ under the action of the wetting interaction term~\eqref{eq:disjoin-p}. The height $h(t)$ is extracted from the mesh, defining the droplet's interface, as the vertical distance between its highest and lowest nodes (see Fig.~\ref{fig:choreo}) over a time interval equivalent to 50 droplet's characteristic times, $t_\mu = \mu_{\rm in} R_0/\sigma$.
To check convergence to the inertia-free BEM implementation, we approach the Stokes' limit by increasing viscosities so that inertia is suppressed. 
This is done by setting $\taulb^{\text{out,in}} = [0.56,0.59,0.732,1.1]$ ($\lambda=1$), leading to decreasing values of the Reynolds number $\Rey$ (see Fig.~\ref{fig:IBLB-BEM}), here defined 
by taking the capillary speed $\sigma/\mu$ as reference velocity and $R_0$ as reference length~\cite{carlson2011dissipation}, hence $\Rey:=\rho_{\rm in}\sigma R_0/\mu_{\rm in}^2$. 
For both solvers, the ratio between the characteristic length $\xi$ and the initial radius $R_0$ of the corresponding spherical droplet is $\xi/R_0=0.05$ and $\rho_{\rm in} = 1$ in numerical units. 
The LB domain considered consists of $250\times250\times150$ lattice nodes, while the Lagrangian nodes used to track the droplet's interface motion are arranged in a mesh of $N_t=120\,000$ triangular faces. Henceforth, we take the lattice spacing and timestep to be $\Delta x=\Delta t=1$, hence dropping them whenever lengths and times are reported. All tests presented in this study were conducted at a fixed capillary number $\Ca$.
Figure~\ref{fig:IBLB-BEM} shows that $h(t)$ extracted from the inertia-free BEM data decreases at a nearly constant rate until $t/t_\mu\simeq10$, then flattens toward a steady state for which $h/R_0\simeq0.96$. A good convergence to the BEM trend is achieved in IBLB simulations by increasing viscosities, thus obtaining a similar profile and steady-state value as $\Rey\to0$. 
This convergence hints at a correct recovery of the hydrodynamic behaviour of our IBLB method in the low-inertia regime. 
However, when inertia is predominant in the IBLB setup, a marked discrepancy is observed from the BEM data: in the IBLB simulations, $h(t)$ initially decreases more slowly than in viscous-dominated cases until a fast drop below the steady-state height value is observed, together with an oscillation becoming more pronounced as $\Rey$ increases. While a strict characterization of this oscillatory behaviour is beyond the scope of the present work, its presence reveals a non-trivial role played by inertia in the observed dynamics.
Next, we therefore consider independent simulations run with a volume of fluid (VoF) solver to further test our method in different setups, including cases where inertia plays a significant role.
\section{Comparison with Basilisk}\label{sec:Basilisk}
As a second test for the hydrodynamic behaviour of our IBLB method, we ran independent simulations using the popular, sharp-interface software \emph{Basilisk}~\cite{popinet2003gerris,popinet2015quadtree}.
Basilisk is an open-source code capable of solving multiphase flows on adaptive Cartesian meshes. Specifically for the simulation of wetting scenarios~\cite{demou2024hybrid,demou2025physics}, it uses (i) an incompressible, two-phase Navier–Stokes solver~\cite{popinet2003gerris}, (ii) a geometric VoF method to capture the movement of the fluid-fluid interface~\cite{popinet2009accurate}, (iii) the continuous surface force method for the surface tension calculations~\cite{popinet2018numerical}, and (iv) height functions to impose the local contact angle on the wall~\cite{afkhami2008height}. 
The adaptive mesh refinement procedure increases the spatial resolution at the fluid-fluid interface and reduces it in less active regions, facilitating efficient and accurate simulations. 

\begin{figure}[t!]
    \centering
    \includegraphics[width=\linewidth]{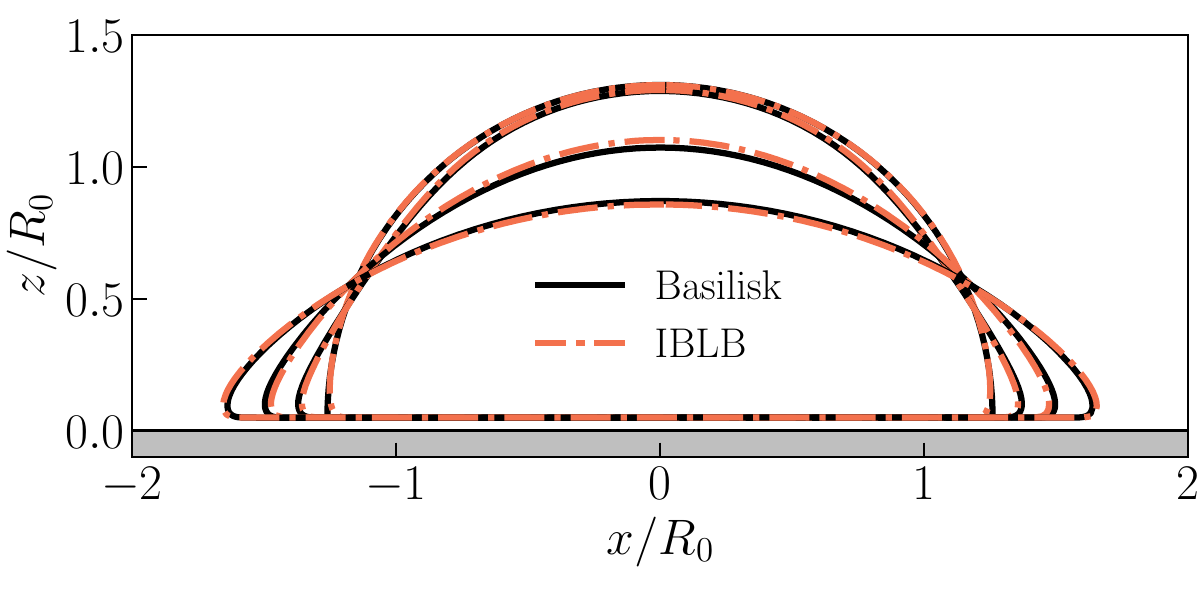}
    \caption{Comparison of the droplet profiles for a spreading process using the IBLB method (dashed lines) and Basilisk (solid lines). The droplet is initialized as a spherical cap (see text for more details) and half-profiles are plotted for $t/t_\mu=[0,4,10,70]$. 
    } \vspace*{-2mm}
    \label{fig:IBLB-basilisk_cap}
\end{figure}

We first check the consistency of the surface tension implementation between the two solvers by comparing the deformation of a free droplet in a shear flow~\cite{pelusi2023sharp}. This is done by placing an initially spherical particle at the center of a channel with a fully developed shear flow.
The deformation is measured via the deformation index $D$, defined as
\begin{equation}
    D(t):=\dfrac{r_1(t)-r_3(t)}{r_1(t)+r_3(t)}\,,
\end{equation}
with $r_1$ and $r_3$ droplet semi-axes in the shear plane.
\begin{figure}[t!]
    \centering
    \includegraphics[width=\linewidth]{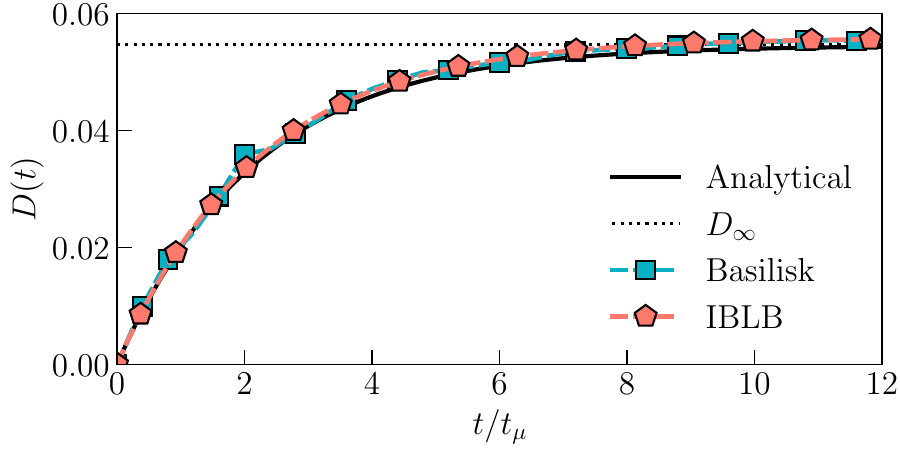}
    \caption{Evolution of the deformation index $D$ for a free droplet under shear. The solid line indicates the analytical solution (cf.~\cite{barthesbiesel1985,pelusi2023sharp}), the dotted line marks the asymptotic deformation value $D_\infty$, pentagons and squares refer to IBLB and Basilisk data, respectively.} \vspace*{-2mm}
    \label{fig:shear_deformation}
\end{figure}
Numerical solutions for the transient phase of the evolution of $D$ can be found assuming small deformations and employing discrete integration techniques, while its asymptotic value $D_\infty$ can be analytically derived as~\cite{barthesbiesel1985,pelusi2023sharp}:
\begin{equation}\label{eq:D_inf}
    D_\infty = \frac{19\lambda+16}{16\lambda+16}\Ca \, .
\end{equation}
This result holds in the low-inertia limit. To match it, we therefore consider a case with high viscous forces (i.e., small values of $\Rey$), and small deformations (i.e., small values of $\Ca$).
Specifically, we employ
$\Rey=0.01$, $\Ca=0.05$, $R_0=19$, and $\mu_{\rm in}=1/6$ (i.e., $\taulb^{\rm in}=1$) in both the IBLB and Basilisk setups.
To achieve the same $\Rey$ and $\Ca$, we set $\rho_{\rm in}=1$ in the IBLB simulations and a shear rate $\dot{\gamma}=1$ and $\sigma=63.3$
in the Basilisk ones so that $\rho_{\rm in}$ is fixed as a result of the previous choices.
As can be seen by Fig.~\ref{fig:shear_deformation}, both solvers achieve good agreement for the droplet deformation both in the transient phase, matching the analytical solution, and in the steady-state, with a relative distance from the asymptotic value of $2.7\%$ for the IBLB and $1.3\%$ for Basilisk, both of which could likely be further reduced by taking smaller values of $\Rey$ and $\Ca$~\cite{barthesbiesel1985,pelusi2023sharp}. However, in the early stages, Basilisk is prone to slight oscillations, most notably around $t/t_\mu=2$, which is probably due to its grid-refinement algorithm. Overall, this comparison shows substantial correspondence between the two solvers' surface tension implementations. 
Having excluded possible discrepancies in the droplet interfacial properties reproduced by the two methods, we set out to assess the agreement between Basilisk and the IBLB for wetting dynamics. 
Preliminary tests, run against the Basilisk setup described at the beginning of the section, showed fundamental inconsistencies with the IBLB simulations (data not reported).
In order to achieve a compelling comparison, we therefore discard the height functions implementation from the Basilisk setup described at the beginning of this section and supply the VoF solver with the wetting interaction term of \eqref{eq:disjoin-p}, resulting in the formation of a thin film separating the droplet from the wall. Our motivation for doing so is twofold. First, the absence of an explicitly imposed slip model and the interaction term $\Pi$ induces contact line motion driven by numerical slip effects~\cite{Afkhami_2018}; in such cases, the spreading dynamics is resolution-dependent and, for all cases considered, typically occurred over shorter time scales than in our model, which presents an extra pressure near the contact line that effectively inhibits the spreading process (see Fig.~\ref{fig:choreo}(b)). Second, the final equilibrium predicted by our model differs from the spherical-cap solution when the interaction term is absent for the values of $\xi/R_0$ considered. Therefore, incorporating $\Pi$ in Basilisk was a necessary prerequisite for a fair comparison between the two solvers. 
Having implemented the same interaction term, we check the agreement between Basilisk and the IBLB method in one case previously considered for the comparison with BEM.
With this aim, we initialize the simulations taking an equilibrated droplet for the IBLB method and a hemispherical droplet of the same volume for Basilisk (see highest-sitting profiles in Fig.~\ref{fig:IBLB-basilisk_cap}). Then, we let them spread towards an equilibrium contact angle $\theta_{\rm eq}=\pi/4$ (lowest-sitting profiles in Fig.~\ref{fig:IBLB-basilisk_cap}). 
For these simulations, we set $\rho_{\rm in}=\lambda=1$, $\mu_{\rm in} = 0.03$ (i.e., $\taulb^{\rm in}=0.59$), and $\xi/R_0=0.05$ in both solvers. We further set $R_0=60$, $\sigma=0.001$ in the IBLB system, and $R_0=1$ and $\sigma=0.06$ in the Basilisk one, so that the codes reproduce the same $\Rey$, namely $\Rey=66.7$.
The IBLB simulation domain is the same as in the BEM comparison section, while the Basilisk simulation is performed in a cubic domain of length $L_0=2.5$ reproducing a single octant of a sphere: the full shape is reconstructed exploiting the axisymmetry of the system.
The agreement between the shapes is found to be remarkable, with a relative distance between the curves (evaluated as the same L2-norm of~\cite{bellantoni2025immersed}) peaking at 1.1\%. This result is non-trivial and represents a significant improvement with respect to~\cite{bellantoni2025immersed}, where the control on the droplet profile was limited to the equilibrium configuration. Here, we instead query and match shapes at arbitrary stages in the spreading process.
Finally, the Basilisk-$\Pi$ implementation allows to compare the two solvers in the inertia-dominated regime, where a ``neck'' reminiscing that of coalescing droplets forms between the droplet and the wall~\cite{bonn2009wetting}. For this comparison, we initialize the droplet as a sphere of radius $R_0$ placed at a distance $d\simeq\varepsilon$ from the wall (see Fig.~\ref{fig:choreo}(c)) and set $\rho_{\rm in}=\lambda=1$, $\mu_{\rm in} = 0.02$ (i.e., $\taulb^{\rm in}=0.56$), $\theta_{\rm eq}=\pi/4$, and $\xi/R_0=0.05$ in both solvers. The reference radius and surface tension for both solvers are the same as in the previous case.
The IBLB simulation was run using a $300\times300\times250$ Eulerian lattice and $N_t\!=\!120\,000$ triangles for the droplet interface, while in the Basilisk simulation the domain size is the same as in the previous case.
\begin{figure}[t!]
    \centering
    \includegraphics[width=\linewidth]{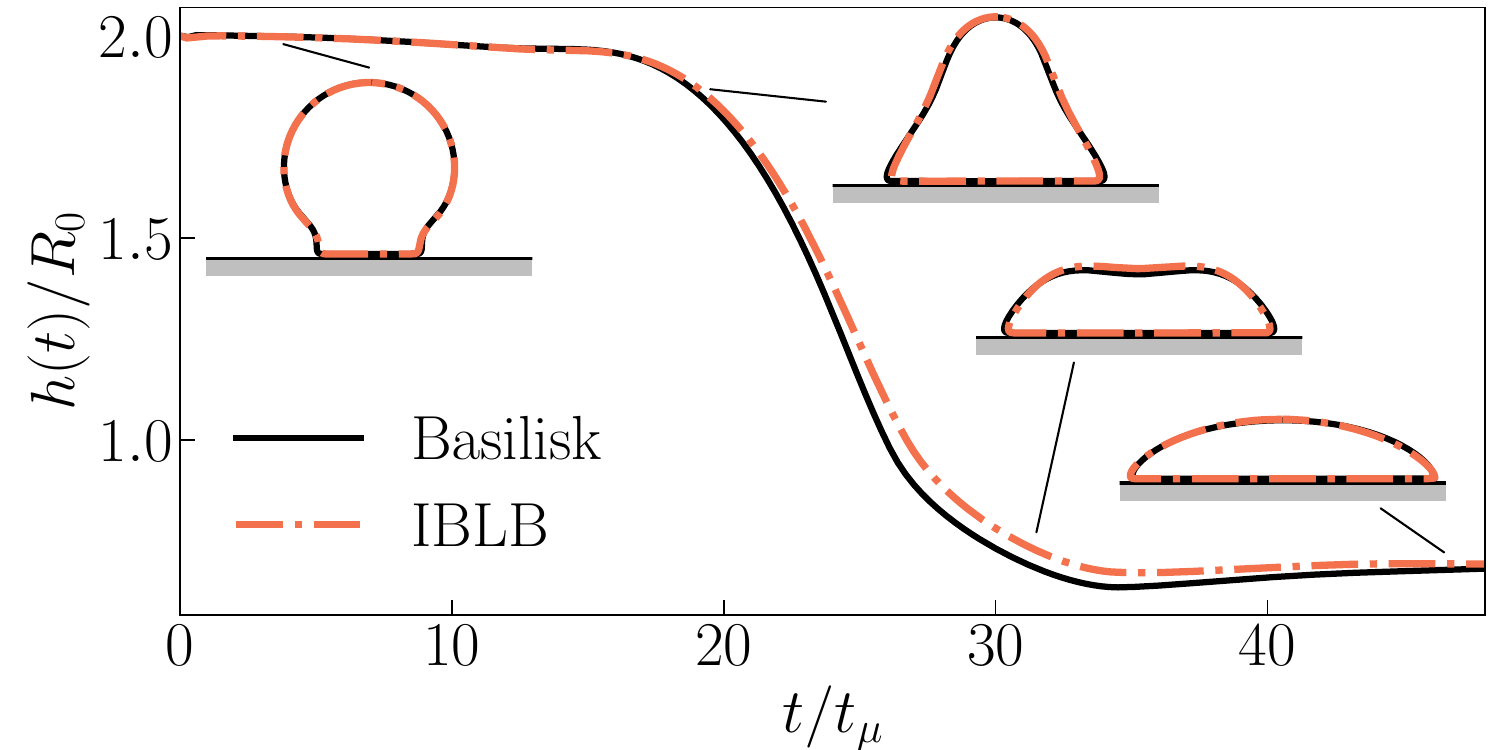}
    \caption{Comparison of height evolution of a spherical droplet using the IBLB method (dashed line) and Basilisk (solid line). Insets show droplet's profiles obtained with Basilisk and the IBLB at $t/t_\mu=[3,18,30,48]$.} \vspace*{-2mm}
    \label{fig:IBLB-basilisk_sphere}
\end{figure}
For this comparison, we consider both the droplet's height evolution and droplet profiles at fixed times (see Fig.~\ref{fig:IBLB-basilisk_sphere}).
Results highlight a good agreement between the two solvers despite a relative error, taken as the relative height difference between the curves, peaking at $9.0\%$ around $t/t_\mu \simeq 25.5$. Furthermore, the methods show the same bouncing effect in the droplet profile, highlighted in the profiles shown as insets of Fig.~\ref{fig:IBLB-basilisk_sphere}, where the formation of a neck in the proximity of the contact line is followed by a fast flattening of the droplet to a ``pudding-like'' shape, which eventually relaxes to equilibrium resembling a spherical cap.
This phenomenon is essentially linked to the presence of inertia in the system. Indeed, the oscillations in the height profile of the curves with the highest $\Rey$ in Fig.~\ref{fig:IBLB-BEM} can be associated to a similar bouncing.
These independent comparisons with Basilisk highlight that, both with and without inertia, the IBLB method is able to capture the shape of the droplet with a contact line model throughout the spreading process, on top of consistently recovering the same hydrodynamics, as attested by the small differences found in the time evolutions of the droplet's height.
\section{Conclusions and Perspectives}\label{sec:conclusions}
In this work, the hydrodynamic behaviour of an IBLB method for wetting dynamics was studied as an extension to the model validation presented in Ref.~\cite{bellantoni2025immersed}.
In particular, focus was placed on the contact line model implemented and its effects at the fluid-solid interface during droplet spreading on a flat homogeneous solid wall.
The interaction between the droplet and the wall was enforced via an interaction term resembling a disjoining pressure. Due to the attractive-repulsive nature of this interaction, a thin film is formed at the interface between the solid and the droplet. 
In order to assess the correct hydrodynamics recovery in the presence of the thin film, we compared the IBLB simulations against two independent solvers: (i) a Stokes' flow solver relying on the boundary element method (BEM), which was implemented for this purpose, and (ii) the volume of fluid solver Basilisk, which solves for multi-phase incompressible NSE.
To achieve meaningful comparisons, both alternative solvers were equipped with the IBLB wetting interaction term (see Eq.~\eqref{eq:disjoin-p}).
Since the droplet dynamics is coupled to the system's hydrodynamics, the droplet's height, deformation, and shape evolution were treated as hydrodynamic probes to analyze the IBLB results against the other solvers.
The comparison with the first method showed that the droplet's height evolution obtained with the IBLB method converges to that achieved with the BEM as the low-inertia limit is approached.
As a full NSE solver, Basilisk offered more flexibility to perform comparisons, since it allows to: (i) independently test the surface tension implementation by taking into account a free droplet under shear, (ii) further examine wetting in the low-inertia limit explored with the BEM, and (iii) compare solutions for early stages of inertia-dominated wetting.
All the cases we examined indicate a correct recovery of the hydrodynamics by the IBLB, with relative errors between the solvers and the analytical results below 10\% in all tests considered. It must be noted that these errors are sensitive to the used computational domain, and better agreement is most likely to be found if the domain size is increased.

In Ref.~\cite{bellantoni2025immersed}, the hydrodynamic behaviour of the method was tested by considering the exponents of known scaling laws for the contact radius dynamics, while the present work critically strengthens the previous method's validation by addressing its hydrodynamic behaviour using one-to-one comparisons of droplets' shapes, heights, and deformation. 
Crucially, the comparisons presented in the current contribution underline that the thin film emerging from our contact line model does not jeopardize the IBLB capability to correctly recover the system's hydrodynamics.
Furthermore, in Ref.~\cite{bellantoni2025immersed}, the droplet shape was examined directly only at steady-state, where a theoretical solution could be derived. The present contribution expands on that by providing a check on the droplet shape during spreading. In this regard, it is significant that our method and Basilisk capture the same ``bouncing'' effect of the droplet upon high-inertia spreading, when the intermediate shapes assumed by the droplet are far from the equilibrium shape.

As an interesting extension to the present work, the contact line model introduced in our IBLB method opens up the possibility of modeling complex flows exhibited by soft particle pastes and gels~\cite{meeker2004slip_a,meeker2004slip_b}.
Additionally, the method's ability to produce correct droplet shapes during spreading potentially enables the study of wetting for particles with complex interfacial properties, such as elasticity, a little-explored topic that has recently attracted interest~\cite{rao2021elastic,pelusi2023sharp,mitra2026}. 
Finally, the control of droplet shapes could be explored in data-driven applications that leverage IBLB data to develop shape-prediction tools with broad applications in microfluidics and materials design.

\acknowledgments
This research is supported by European Union’s HORIZON MSCA Doctoral Networks programme, under Grant Agreement No.\,101072344, project AQTIVATE (Advanced computing, QuanTum algorIthms and data-driVen Approaches for science, Technology and Engineering). FP acknowledges the Innovative Medicines Initiative 2 Joint Undertaking (JU), grant agreement No 853989. The JU receives support from the European Union’s Horizon 2020 Research and Innovation Programme and EFPIA and Global Alliance for TB Drug Development Non-Profit Organisation, Bill \& Melinda Gates Foundation, University of Dundee.
FG acknowledges the Italian Ministry of University and Research (MUR) under the FARE program (No. R2045J8XAW), project ``Smart-HEART''. Support/funding from Tor Vergata University project AI4HEART and INFN/FIELDTURB project are also acknowledged.
MD acknowledges partial funding from an Inria Chair. 
\bibliographystyle{eplbib}
\bibliography{refs}
\end{document}